\documentclass{article}
\usepackage{icrctc07}

\newcommand{\gray}[1]{$\gamma$-ray{#1}}

\newcommand{\pubjournal}[6] {#1, #2 {\bf #3}, #4 (#5).}

\title{Isotopic composition of cosmic-ray sources}

\shorttitle{Isotopic composition of cosmic-ray sources}

\abstract{
We use the GALPROP code and the Advanced Composition Explorer (ACE) 
data to derive the cosmic ray (CR)
isotopic composition at the sources. 
The composition is derived
for two propagation models, diffusive reacceleration and plain
diffusion. 
We show that the compositions derived assuming
these two propagation models are different.
We also compare the isotopic composition at the sources with
the latest solar composition.
}

\authors{I. V. Moskalenko$^{1,2}$, A. W. Strong$^{3}$, and T. A. Porter$^{4}$ }
\shortauthors{Moskalenko and Strong}
\afiliations{
$^1$Hansen Experimental Physics Laboratory, Stanford University, Stanford, CA 94305, U.S.A.\\ 
$^2$Kavli Institut for Particle Astrophysics and Cosmology, Stanford University, CA 94309, U.S.A.\\
$^3$Max-Planck Institut f\"ur extraterrestrische Physik, Garching, Germany\\
$^4$Santa Cruz Institute for Particle Physics, University of California, Santa Cruz, CA 95064, U.S.A.}
\email{imos@stanford.edu}

\begin{document}
\maketitle

\section{Introduction}
CR source abundances are normally derived using the leaky-box or
weighted slab approximations to interstellar propagation.  While this
is a valid procedure under some conditions (see \cite{AnnRev57} for a
review) at least for stable nuclei, there are reasons for preferring a
more physically-based approach.   For example, the distribution of CR
sources (e.g., supernova remnants) in the Galaxy is probably peaked
towards the inner Galaxy, as recently supported by \gray{}
observations \cite{SMRRD04}, and this affects the path-length
distribution in a way that requires a detailed spatial propagation
model.  Another example is the effect of diffusive reacceleration,
which is probably important in reproducing the energy-dependence of
the secondary/primary ratios, and this will also affect the derivation
of source abundances.   The parameters of the models used to derive
the source abundances should also be compatible with other
observational constraints from gas surveys, \gray{s}, synchrotron and
so on  \cite{SMR00,Moskalenko2003,SMR04,SMRRD04}.

In this paper, we use the GALPROP CR propagation code \cite{SM98} to
derive source isotopic abundances from ACE data.  Both the
astrophysical model and the cross-sections play a key role in the
uncertainties in such a computation; here we present preliminary
results, reserving detailed discussion to a future journal paper.

\section{Method}
The GALPROP code\footnote{http://galprop.stanford.edu}
computes a complete network of primary, secondary and tertiary
production starting from input source abundances, 
as described in \cite{SM98,SMR04,Ptuskin06}.
The nuclear reaction network is built using the
Nuclear Data Sheets. The isotopic cross-section database
is built using the extensive T16 Los Alamos compilation
of the cross-sections \cite{Mashnik1998} and modern
nuclear codes CEM2k and LAQGSM \cite{Mashnik2004}. 
The most important isotopic production cross-sections
(2H, 3H, 3He, Li, Be, B, Al, Cl, Sc, Ti, V, and Mn)
are calculated using our fits to major production channels
\cite{Moskalenko2001,Moskalenko2003}. 
Other cross-sections are calculated using
phenomenological approximations by Webber et al. \cite{Webber1990}
(code WNEWTR.FOR versions of 1993 and 2003)
and/or Silberberg and Tsao \cite{ST1998} (code
YIELDX\_011000.FOR version of 2000) renormalized to
the data where it exists. The
K-capture and electron stripping processes are included,
where a nucleus with one electron is considered a separate 
species because of the difference in lifetime. 


\begin{figure*}[t]
\centerline{\includegraphics[width=0.95\textwidth]{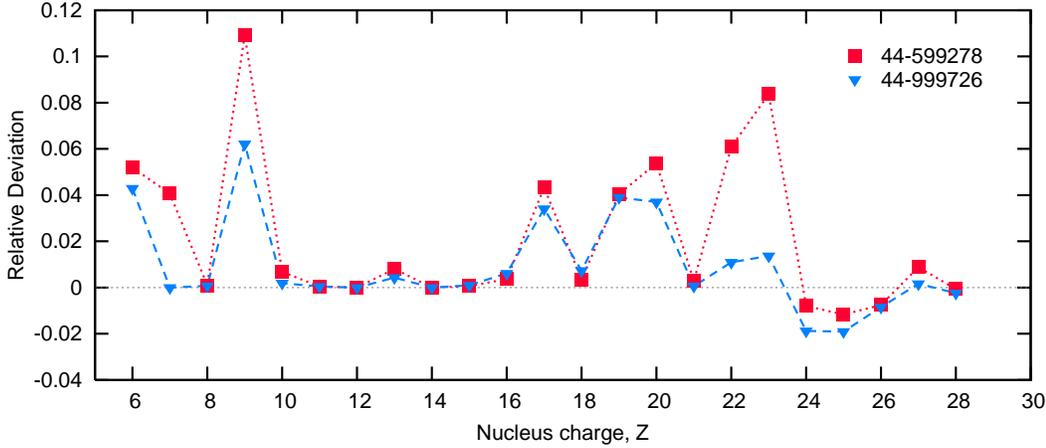}}
\caption{\label{fig:propagated_abundances}
Quality of the fit: fractional deviations of 
propagated elemental abundances from ACE observations \cite{Wiedenbeck2001}.}
\end{figure*}

The propagation equation is solved numerically starting with the
heaviest nucleus (i.e., $^{64}$Ni), computing all the resulting
secondary source functions, and then processes the nuclei with $A-1$.
The procedure is repeated down to $A=1$.  To account for some special
$\beta^-$-decay cases (e.g., $^{10}$Be$\to^{10}$B) the whole loop is
repeated twice. The current version employs a full three-dimensional
spatial grid for all CR species, but for the purposes of this study
the 2D cylindrically symmetrical option is sufficient.

To calculate the isotopic source abundance, we adopt an iterative
procedure, which uses the deviations of calculated propagated
abundances from abundances observed by ACE at 200 MeV/nucleon
\cite{Wiedenbeck2001} to correct the source abundances until a good
fit is obtained.  In practice very good agreement (a few \%) can be
obtained in about ten iterations.   The basic parameters of the
propagation model are based on fitting the energy-dependence of the
B/C ratio, for which the best data are available and cross-sections
best known.  The parameters are given in \cite{Ptuskin06}.  Solar
modulation is calculated using the force-field approximation
\cite{forcefield}.  The modulation potential  $\Phi=450$ MV
corresponds approximately to the period of solar activity when the
data were collected.  An analysis of source  abundances of isotopes of
P, S, Ar, Ca is given in \cite{Ogliore2007}.  Discussion of
propagation of Li, Be, B in the same models can be found in
\cite{deNolfo2006}.

\begin{figure*}[t]
\centerline{
\includegraphics[width=0.96\textwidth]{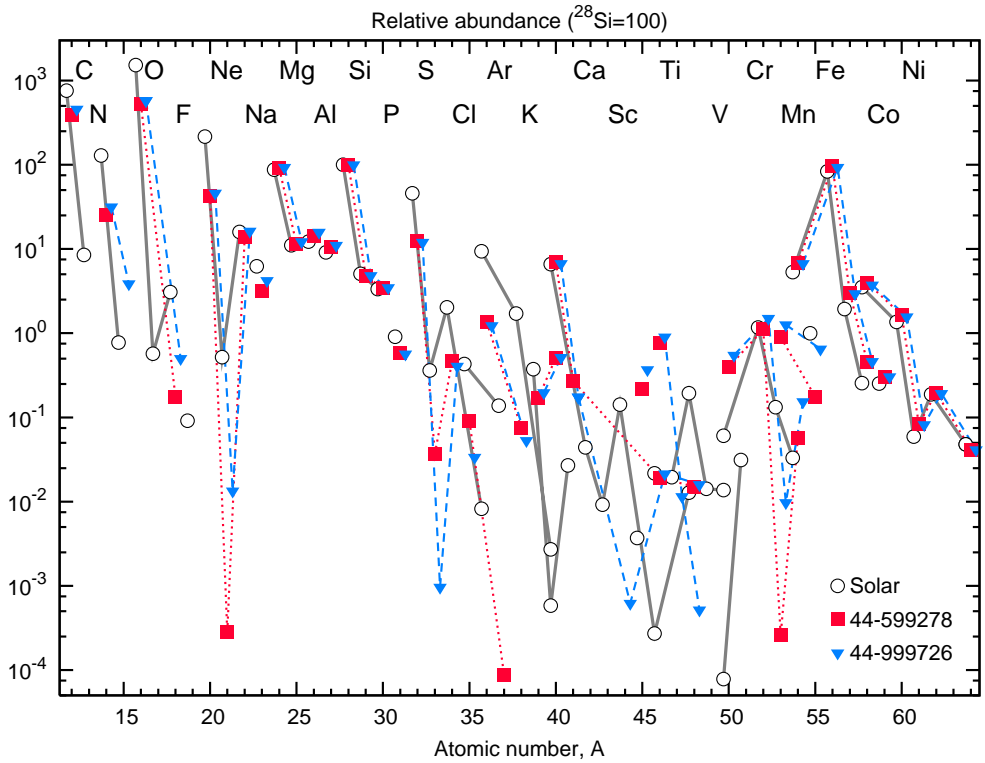}}
\caption{\label{fig:source_abundances} 
Source isotopic abundances.
Circles: solar 
system abundances \cite{iso_solar},
squares: diffusive reacceleration model,
triangles:  plain diffusion model.
Isotopes of the same element are connected by the lines.
The lowest values are included to show the solar abundances;
in these cases the CR source abundances are not always reliable.
}
\end{figure*}

\section{Results}
We applied the technique to the plain diffusion and the diffusive
reacceleration models as described in \cite{Ptuskin06}.  We consider
here only the effect of the propagation; the effect of a more
realistic source distribution will be considered elsewhere (Moskalenko
et al., in preparation).  The quality of the fit can be judged from
Figure~\ref{fig:propagated_abundances}, which shows the propagated
abundances of elements minus abundances measured by ACE.  The fitted
values are \emph{isotopic} abundances while the plot shows fractional
deviations of  \emph{elemental} abundances for clarity.  It can be
seen that the propagated abundances  are reproduced with a maximum
error of 5\% (10\% for the least  abundant elements).

The deviations from ``0'' are mainly due to the errors in the cross
sections.  It can be illustrated using Carbon as an example.  The
propagated abundance of $^{12}$C agrees perfectly with the data since
the source abundance of this isotope is adjusted iteratively.  The
isotopic abundance of $^{13}$C cannot be adjusted in the same way
since it is already zero (from the fitting), but it is overproduced
due to the spallation of heavier nuclei (see discussion in
\cite{Moskalenko2003}).  The entire excess of Carbon $\sim$5\% is due
to the overproduction of $^{13}$C during the CR propagation.
Similarly, F is entirely secondary and the excess is due to the
cross-section uncertainties.

The isotopic abundances ($A>6$) for the two models are compared with
latest solar abundances from \cite{iso_solar} in
Figure~\ref{fig:source_abundances}.  We do not show the isotopes where
the source abundances are extremely small ($<$$10^{-4}$) because the
accuracy of the cross sections is insufficient in these cases.
The derived source isotopic abundances generally agree better with the
latest solar abundances from \cite{iso_solar},  especially the iron
group, than with the  earlier version of solar abundances by Anders
and Grevesse \cite{AG1989}.  Still, the source abundances derived in
both models are underabundant relative to solar in many cases, as is
well known, e.g.\ C, N, O, Ne, S, Cl, Ar, Sc.  In other  cases we
confirm the remarkable agreement with solar:  Na, $^{40}$Ca, Mg, Al,
Si,  $^{52}$Cr, Fe, Co, Ni.  In many cases a good agreement with solar
abundances is apparent also for isotopes with large proportion of
secondary production: $^{22}$Ne, $^{23}$Na, $^{25,26}$Mg, $^{27}$Al,
$^{29,30}$Si, $^{31}$P, $^{54,57,58}$Fe, $^{59}$Co, $^{61}$Ni.

The abundances are very much dependent on the isotopic  production
cross sections which are uncertain to a large degree, e.g., F, P, Ar,
K, Sc, Ti, V (see discussion in \cite{Moskalenko2003}).  The source
abundance of $^{16}$O, which is mostly primary, is less than solar by
a significant factor.  The fragmentation cross sections are known much
better than the isotopic production cross sections, so this result is
rather robust.  Similarly, we confirm that the source abundances of
$^{14}$N and $^{20}$Ne  are a factor of 5--6 below solar.  The
well-known excess $^{22}$Ne/$^{20}$Ne  (see, e.g., \cite{Binns05}) is
also evident.  Subject to further verification and the accuracy of the
production cross sections, the study shows that many isotopes in CR
are mostly secondary, $^{13}$C, $^{17}$O, $^{21}$Ne, $^{33,34,36}$S,
$^{37}$Cl, $^{38,40}$Ar, $^{41}$K, $^{42,43,44}$Ca, $^{53}$Cr.  This
is in addition to Li, Be, B, F, P, Sc, Ti, V that have been discussed
in the literature for a long time.
There are also a few cases where the  model predictions are different
from each other, e.g.,  $^{15}$N, $^{18}$O, $^{21}$Ne, $^{33}$S,
$^{55}$Mn, and some others.  These differences can be used to
constrain or rule out  some propagation models.

Interestingly, radioactive $^{41}$Ca and $^{53}$Mn appear to be
significantly present in the sources.  Radioactive $^{41}$Ca is a
K-capture isotope that decays (to $^{41}$K) with a period
$\tau_{1/2}=1.03\times10^5$ yr and  has a negligible abundance in the
solar system.  A similar case is  radioactive $^{53}$Mn, another
K-capture isotope that decays to $^{53}$Cr with a period
$\tau_{1/2}=3.74\times10^6$ yr.  Their abundances can be used to
constrain the CR acceleration time scale.


\section{Conclusion}

This is the first time that a `realistic' (i.e.\ full spatial- and energy-dependence)
propagation model has been used to derive isotopic source abundances for a 
full range of CR nuclei. As is well-known, 
the elements with low first ionization potential (FIP) appear to be more
abundant in CR sources relative to the high-FIP elements, when compared
with the solar system material. This might imply that the source material
includes the atmospheres of stars with temperatures $\sim$$10^4$
K \cite{casse78}. A strong correlation between FIP and volatility (most
of low-FIP elements are refractory while high-FIP elements are volatile)
suggests that CRs may also originate in the interstellar dust, pre-accelerated
by shock waves \cite{meyer97,epstein80}. CR data tend to prefer volatility
over FIP, but uncertainties in the derived source abundances have prevented an 
unambiguous solution. 

The analysis presented in this paper shows 
that the radioactive $^{41}$Ca and $^{53}$Mn
appear to be significantly present in the sources. This may be used
to study the CR acceleration timescale and
seems to give more support to the volatility hypothesis.
The detailed discussion and error analysis will be presented in a 
forthcoming journal publication.



{\bf Acknowledgements.}
I.\ V.\ M.\ acknowledges partial support from NASA APRA
grant. T.\ A.\ P.\ is supported in part by the US Department of Energy.

\end{document}